\documentstyle[12pt,aasms]{article}
\slugcomment{AJ, revised manuscript, LaTeX, postscript figures}

\begin{document}

\title{Field Distortions in the CPC2 Plate Data}
\author{N. Zacharias\altaffilmark{1}}
\affil{U.S. Naval Observatory AD, 3450 Mass. Av. N.W., Washington D.C. 20392}

\altaffiltext{1}{with University Space Research Association (USRA),
  Division of Astronomy and Space Physics, Washington D.C.}

\begin{abstract}
Residuals of a conventional plate adjustment of 5,687 plates
of the Second Cape Photographic Catalogue (CPC2)
using the Southern Reference Star (SRS) Catalog
were analyzed to search for systematic errors depending on
the location of the star images on the plate.
Even after correcting for third order optical distortion,
such coordinate depending systematic errors,
hereafter called field distortion pattern (FDP),
were detected which are nearly constant with respect to
plate epoch, location in the sky and apparent brightness of the stars.
The mean amplitude of these field distortions was found to be
57 milliarcseconds (mas), which is $0.57 \ \mu m$ on a plate.
Due to the favorable 4--fold overlap observing principle of the CPC2,
the effect on the mean catalog position is only a fraction of that.
But the FDP affects a block adjustment (BA)
of the entire set of data on a 100 mas level at the border of the
CPC2 sky coverage.
A comparison with FDP's found in other projects was made
and physical reasons for the FDP were investigated.
Publication of a new release of the CPC2 is planned after
the Hipparcos Output Catalog becomes available in 1996.
\end{abstract}


\section{Introduction}

The Second Cape Photographic Catalogue (CPC2)
is a major astrometric catalog of high
star density in the southern hemisphere and has contributed significantly
to compilation catalogs such as the ACRS (Corbin \& Urban 1992)
and the PPM (Roeser \& Bastian 1993). All derived proper motions
for fainter stars in the southern hemisphere are influenced by the
precision and systematic accuracy of the CPC2.

The CPC2 covers the entire southern sky
south of $+2\deg$ declination. Positions of 276,131 stars to a limiting
magnitude of $V \approx 10.5$ at a mean epoch of $\approx 1968$ were
published in Paper I (de Vegt et al. 1993).
A precision of 60 milliarcsecond (mas) at the epoch of observation
was obtained for the $\approx 250,000$ quality class 1 and 2 stars.
The catalog is based on 5,687 plates taken at the Cape Observatory
in South Africa from 1962--72.
 All plates were measured on the GALAXY machine
at the Royal Greenwich Observatory (RGO) in the UK.
The published first release of the CPC2 is based on a conventional
plate adjustment, a joint effort of RGO and Hamburg Observatory.
The Southern Reference Star Catalog (SRS) (Smith et al. 1990)
was used for separate reductions of the star positions to the
FK4 and FK5 systems.

A detailed description of the CPC2 project and availability of the
first release of the catalog is given in Paper I, while details of the
reduction procedure can be found in Paper II (Zacharias et al. 1992).
The second release of the CPC2 will be based on a rigorous block adjustment
(BA) of the entire plate data (Zacharias \& de Vegt 1996, Paper III).
In order to find the optimal physical plate model and fully investigate
possible systematic errors in the photographic data,
the Hipparcos Output Catalog will be used first for a re--investigation
of the conventional plate adjustment process.
Because the Hipparcos Output Catalog will soon become available,
there are no plans
to publish an intermediate CPC2 release based on a block adjustment
using the currently available transit circle catalogs.

This paper is intended as a research note towards the final reduction
of the CPC2.
Section 2 will describe the observational data and peculiarities
relevant to the investigations made here.
Section 3 briefly describes the reduction methods.
Because of the large amount of data available in the CPC2 project,
significant results can be obtained also from subsets of the data.
Thus, the stability of the field distortions and their dependence on time
(plate epoch, season), location in the sky and magnitude will be
investigated in Section 4.
The improvement of the individual and mean star positions by applying
the FDP is estimated in Section 5.
Problems and their solutions arising from the block adjustment of the
CPC2 plate material will be investigated in Section 6.
Finally, in Section 7 a discussion of the physical reasons for these FDP's
and comparisons with other projects is made.

\section{Observations}

All plates were taken with the Cape Astrograph of
2--meter focal length in a yellow--red spectral bandpass (530--640 nm)
in a 4--fold overlap pattern.
The measured plate area is $ 4.07 \deg \times 4.07 \deg$ with a plate scale
of $100"/mm$.
Observations were made in 3 groups.
All plates of the Cape Zone with plate centers
in the declination range $-52 \deg \le \delta \le -40 \deg$
were taken first between 1961.5 and 1962.9,
followed by the zone $-40 \deg \le \delta \le -30 \deg$
observed between 1966.5 and 1967.9, and all remaining
southern hemisphere plates (equatorial and polar region
quasi simultaneously) were taken between 1968.2 and 1972.0,
with the exception of 3 plates taken in 1974.

The camera was mounted on the Victoria telescope for the
Cape Zone observations and then moved to the
Multiple Refractor Mount (MRM) for all other observations
of the CPC2 program.
The camera was rotated by $90\deg$
with respect to the sky in this relocation process.
The same focal plane filter and plate holder was used throughout
the entire observing period and the camera was always on
one side of the pier at both telescope mountings.

There are 2 exposures on each plate with equal exposure times of usually
3 minutes, with the second exposure taken immediately after the
first and shifted by about $30\arcsec$ \ in declination.
The regular 4--fold overlap pattern with plate centers on
$0\deg, -2\deg, -4\deg ... -90\deg$ thus gives 8 images per star.
All observations were made close to the meridian.
The use of a yellow--red spectral bandpass, as opposed to the blue
bandpass widely used in older photographic astrometric
work, reduces dispersion and refraction effects of the Earth's atmosphere.

All the nearly 6,000 plates were measured
at Royal Greenwich Observatory (RGO) in the UK on the GALAXY astrometric
measuring machine.
First, the Cape Zone plates were measured in 1978--80,
followed by the polar cap plates ($\delta \le -80 \deg$),
followed by the equator zone ($-20 \deg \le \delta \le 0 \deg$),
the ($-78 \deg \le \delta \le -54 \deg$) zone, and finally
the ($-38 \deg \le \delta \le -22 \deg$) zone.
Measures were completed in 1983.

All plates were rotated by $180 \deg$ with respect to the coordinate
system of the measuring machine between 2 sets of measurements.
Combined x,y data of the 2 sets of measures per plate were obtained
by a linear transformation.
These combined data were used for the published
CPC2 catalog as well as for this investigation.
The precision of such a combined measurement of a single image is about
$\sigma_{xy} = 1.1 \mu m = 110 mas$ (Paper II).

For further details about the GALAXY machine and the measuring process,
the reader is refered to Murray \& Nicholson (1975)
and Nicholson (1979).

\section{Reduction procedure}

\subsection{Combining exposures}

In the first reduction step, x,y measures of both exposures on each plate
were combined by a linear transformation using least squares
and eliminating outliers by a $3 \sigma$ criterion.
Because of the observation principles, no non--linear effects are
expected within the short period of time of a few minutes during which
both exposures were taken.
This assumption was verified by results of
the previous conventional plate adjustment (CPA), where similar values
for plate constants were found for the separate reduction
of the individual exposures of each plate (Paper II).

\subsection{Conventional plate adjustment}

Using the combined x,y data of both exposures of each plate,
a new CPA was performed
using the current version of the Hamburg Block Adjustment
Program Package, HBAPP (Zacharias 1987) which was
transfered to the U.S. Naval Observatory (USNO).

A constant third order optical distortion term of $40 \ mas / degree^{3}$
was applied to all x,y data prior to the adjustment.
Mean SRS reference star positions were transformed into
observed places, applying a rigorous apparent place and
refraction routine.
An 8 parameter plate model, with 4 orthogonal, 2 non--orthogonal
linear, and 2 plate tilt terms, was used throughout this
investigation for the least--squares plate adjustment.
Weights of reference stars were calculated using the mean SRS star
position precision of $\sigma_{\alpha cos \delta} = 90 \ mas,
\sigma_{\delta} = 110 \ mas$ at the epoch of SRS observation,
an estimated error in the proper motions of $\sigma_{\mu} = 4.5 \ mas/yr$
and adopting a conservative mean error of
$\sigma_{xy} = 90 \ mas$ for the precision of the photographic image
(combined exposures, single plate).
The mean epoch of the SRS is also $\approx 1968$ thus $\le 6$ years
of epoch difference (CPC2--SRS) had to be bridged for individual
reference stars on individual plates.
On the average, there are about 14 reference stars on a plate.
All plate reductions were performed with the FK5/J2000 version
of the SRS.

\subsection{Constructing the field distortion pattern}

Differences (photographic position -- reference star position)
in $\Delta \alpha_{cos \delta}, \Delta \delta$, as well as
in $\Delta \xi \approx \Delta x,\Delta \eta \approx \Delta y$,
were stored for each reference star on each plate
($\xi, \eta$ are the standard coordinates),
together with auxiliary data such as
magnitude of the star and x,y position of the image on the plate.
The $\Delta \xi, \Delta \eta$ differences were then binned as
functions of the x,y coordinates.
In the discussion to follow, we took $\Delta \xi$ for $\Delta x$
and $\Delta \eta$ for $\Delta y$ because the measuring machine
coordinate system was aligned to the standard coordinate system
with sufficient accuracy.
Outliers were excluded here if $\Delta x, \Delta y \ge 1.0 \arcsec$.
Fewer than 1 \% of the differences were identified as outliers.

For this reduction step several input filter options were
implemented in order to select plates by epoch or location
in the sky, as well as stars by magnitude.
Unweighted means of the differences $\Delta x, \Delta y$
were calculated for all selected differences within each bin and
assigned to the nominal center position of the bin,
thus the field distortion pattern (FDP) was established.

A bin size of $20\arcmin$ was adopted here, resulting in
13 by 13 bins on the plate area.
A total of $79,313$ differences per coordinate
were obtained, an average of $469$ observations per bin
for the vectors of the FDP from all data.
The formal standard error of a component of a single FDP vector
was found to be $\approx 4.7 mas$ by using the scatter of the individual
differences $\Delta x, \Delta y$ which built up the FDP vectors.
Using subsets of the data, this formal standard error is about
a factor of 2 to 3 larger, depending on the mean number of observations
in a bin.

\subsection{Applying the field distortion pattern}

The FDP was applied to the original measures by a simple mask
method in order to obtain corrected x,y measures.
By linear interpolation the $13 \times 13$ FDP grid of all data was doubled
in size along each coordinate, resulting in a $10 \arcmin$ grid size.
The resulting FDP was used directly as a lookup table for
all x,y corrections without any further weighting or smoothing.
This simple scheme is justified by the small values of the
corrections (averaging approximately half a micrometer) and by the already
smooth pattern obtained by the adopted grid size
and the large number of plates in the CPC2 project.

After repeating the CPA with the corrected x,y data, another FDP
was calculated. The remaining small systematic errors were
corrected by one more iteration of the whole process.
The existence of a remaining pattern after the first step can
be explained by the CPA reduction principle. The full amount of
a possible systematic x,y error will not show up in the residuals.
An independent high precision star catalog would be required
to establish the FDP at once instead of using the reference stars from
the CPA.

\subsection{Block adjustment}

A rigorous Block Adjustment (BA) of all 5,687 plates was
performed using the HBAPP (Zacharias 1987).
Observation equations of individual stars with
$\Delta x, \Delta y > 400 mas$
were downweighted by a factor of 2 and
observation equations with $\Delta x, \Delta y > 800 mas$
were eliminated.
Observation equations of stars outside the magnitude range
$6.0 \le V \le 11.0 $ were downweighted by a factor of 10.

As with the CPA, the combined exposure x,y data and the SRS
reference star catalog were used with the same 8 parameter
plate model and the same pre--applied third order optical distortion
term.

Original x,y data and field--distortion--corrected x,y measures
were used in 2 separate BA runs with otherwise identical
parametrizations and weighting schemes.

\section{Results}

\subsection{Field distortions from all data}

The FDP resulting from use of all CPC2 data is displayed in Fig.1.
The maximum length of a vector
(photographic position - reference star position)
is $160 mas = 1.6 \mu m$; the average
vector length is $57 mas$, which was obtained by an
arithmetic mean of the FDP bin values weighted with the number
of observations within a bin.
A test was made with a smaller (10\arcmin) bin size,
but no more details in the structure could be found than are seen
in the FDP of Fig.1.

The pattern is nearly east--west symmetric to the y--axis,
but there is a noticeable north--south asymmetry.
Each quadrant is nearly symmetric with respect to the diagonals
across the plate.
Obviously the corners of the plate are key points in the
symmetry properties of this pattern.
In the plate center the FDP vectors are much smaller than
at the border of the plate.

In order to eliminate possible correlations due to the fact
that the same reference star appears on several overlapping
plates, a FDP was obtained from {\em non}--overlapping plates.
Only every third declination band and only every third
plate within each band were used to obtain that FDP.
The statistic is poorer than on Fig.1, but otherwise
it shows the same pattern.
Here only quantitative results are given in Table 1.

In Fig.2 a and b the radial residuals before and after
correcting for the FDP are shown, respectively.
Obviously applying the FDP has cured this problem, as expected.
Some significant tangential residuals found in some zones
earlier (Paper II) also have vanished after applying the FDP.

\subsection{Field distortions as function of declination zone}

Because of the observing and measuring history, FDP's from subsets as
a function of declination zone should be considered first.
Results from all subsets discussed here are summarized in Table 1.
All patterns look very similar.
The biggest deviation from the mean pattern shown in Fig.1 is found
in the Cape Zone with plate centers in the declination range
$-52\deg \le \delta \le -40\deg$ (Fig.3a).

In order to visualize and quantify the variations of the pattern
as function of the selected subset of the data,
difference vector plots were calculated on a bin--to--bin
basis in the sense (subset pattern $-$ all data pattern).
An example is given in Fig.3b.

The average lengths of the difference vectors are only
$\approx 10 mas = 0.1 \mu m$ on the plate (see Table 1).
This is on the same order as their estimated errors.
A systematic pattern in the difference FDP for the Cape Zone (Fig.3b)
can be seen in the NW and SE corners; otherwise, it is simply noise,
as for the equatorial zone (not shown here).
Thus, the FDP is nearly constant as a function of declination zone.

The FDP for the polar zone (not shown) again is very similar to Fig.1.

\subsection{Field distortions as function of epoch}

The Cape Zone subset of the previous paragraph serves already
as an example for the investigation of the dependence of the FDP on epoch.
All Cape Zone plates were observed within the short period
of $\approx 1962-1963$. All other plates were taken later.

Another example, this time for the latest plate epochs
(1971--1973) in the CPC2 project has been investigated and
quantitative results are found in Table 1.
The FDP again looks very similar to Fig.1 and is not shown here.

\subsection{Field distortions as function of right ascension}

Two subsets as a function of right ascension
$3^{h} \le \alpha \le 6^{h}$ and
$17^{h} \le \alpha \le 20^{h}$ were investigated as well.
The FDP's look very similar to Fig.1 and quantitative results
are listed in Table 1.

\subsection{Field distortions as function of magnitude}

Two subsets as a function of magnitude of the stars were
formed for faint ($V \le 9.0$) and bright ($6.0 \le V \le 7.5$) stars.
The results are listed in Table 1
and the FDP's look similar to Fig.1.

There is an offset in the mean x,y coordinates as a result of
a magnitude equation in declination for the zones north of
$-30\deg$ of about $40 mas/mag$ (see Paper II).
Otherwise there is no significant deviation of the FDP
from subsets as a function of magnitude to the FDP from all data.

\section{Improvement of individual and mean positions}

Each star appears on an average of 4 plates.
It is common practice to use the individual field star positions
on overlapping plates in order to
estimate the precision of an individual star position,
say $\sigma_{\alpha \delta }$, derived from a conventional
plate adjustment.
This method, however, does not represent the best precision
obtainable from the photographic data nor the precision of the
corresponding  measure error, $\sigma_{xy}$.
The value of $\sigma_{\alpha \delta }$ does include contributions
from error propagation of the CPA process due to
plate constant errors, correlations with magnitude and coma
dependent systematic errors, and of course contributions from the FDP.
Rather, $\sigma_{\alpha \delta }$ should be considered as one
statistically available measure in the entire error budget estimation.

Fig.4 shows the mean errors of
$\sigma_{\delta}$ ( $\sigma_{\alpha cos \delta}$ is similar)
of a single observation (single plate but combined exposures)
as a function of the magnitude of the stars, as derived from
the scatter of individual field star positions on overlapping plates.
Two separate CPA's were run, one with the original x,y data and
the other with the x,y data corrected for FDP as described above.
The improvement is consistent with the removal of an error contribution
of $\sigma = 75 mas$ per coordinate from the original data.
This result is nearly independent of the magnitude group.
After the correction of the x,y data for FDP, the positions on
individual plates are closer to their mean catalog positions.
This improvement holds for the {\em individual positions},
but not necessarily for the {\em mean catalog positions},
as will be explained now.

Each star usually appears on 4 plates with its images
systematically shifted according to the FDP.
These images appear in different areas on the
overlapping plates, thus their positions are also shifted
by different amounts and directions.
This increases the scatter of individual positions around
the mean. When applying corrections for the FDP to the data,
this source of scatter is taken out.
If the different shifts by FDP of these 4 images
cancel exactly out, then the same mean position of that star
would result from the raw data as well as from data
after correcting for FDP.
Usually the systematic shifts cancel out only partly,
which will give a somewhat improved mean position
after correcting for FDP.

In order to see how much the {\em mean catalog positions}
were improved by the corrected x,y data,
the effect of the FDP on the 4--fold overlap pattern adopted in
the CPC2 project needs to be calculated.
Assuming a {\em regular} 4-fold overlap pattern (center in corner),
the combined FDP is shown in Fig.5.
Much smaller correction vectors are found here.
Thus, the actual improvement of the mean catalog position of
the stars is only $\sigma_{x} \approx 16 mas,
\sigma_{y} \approx 11 mas$ on average,
as derived from the mean length of vector components of Fig.5.

\section{Implication to block adjustment}

Two separate, rigorous BA's of the entire data were performed
as described above.
The original x,y data were used for one BA, and x,y data
corrected for FDP were used for the other, with an otherwise
identical reduction scheme and parametrization.
The resulting positions were compared to their respective
CPA solutions.
Fig.6 shows the differences (BA--CPA) in $\Delta\delta$
vs. $\delta$ for both solutions plotted to the same scale.
In order to visualize the systematic effect, the data first were
sorted according to declination and then averaged over
1,000 consecutive data points to form the normal points plotted here.

The affect in Fig.6a has disappeared in Fig.6b, where the
corrected x,y data were used.
In the declination range $0\deg \le \delta \le +2\deg$,
only a 2--fold overlap pattern exists, with a 50\% overlap
area along right ascension.
Fig.7 shows the combined FDP of such a 2--fold overlap pattern.
The sign and amplitude of the $\Delta \delta$ systematic
differences on the upper part (north) of Fig.7 are consistent with the
(CPA--BA) results of Fig.6a north of the equator
under the assumption that the CPA positions are close to the system
of the reference stars (reducing the effect of the systematic errors
of the x,y data)
and the BA positions are close to the system of the x,y data including
their systematic errors.
This assumption is justified by the different adjustment
principles of the CPA and BA.
In the CPA the x,y data of only the few reference stars on
a particular plate are used in the adjustment.
Their systematic error due to FDP will be partly compensated
by the reference star catalog positions.
Contrary, all x,y data is used with high weight in the BA,
while the few reference star coordinates will basically provide
only the overall zeropoint for the location in the sky.

\section{Discussion}

\subsection{Comparison with other investigations}

Field distortions were found already on Schmidt plates
(Bucciarelli et al. 1992).
Vector lengths exceeding 1\arcsec or $15 \mu m$ were found there,
an order of magnitude larger than with the CPC2 plates.
The reason for these large distortions is clearly related to
the bending of Schmidt plates in order to match
the curved focal plane.

Plates of the Astrographic Catalog (AC) were investigated
(Abad  1993) using a different approach.
The original measures of the Paris Zone were used by Abad,
together with FK5 stars as reference stars, in a BA.
Vector lengths up to about $350 mas$ were found in the FDP,
corresponding to $6 \mu m$, also much larger than with the CPC2 plate data.

Other AC zones show FDP's related to the reseau lines used for
relative x,y measures up to about $200 mas \approx 3.5 \mu$ after correcting
for various other effects (Urban \& Corbin 1995).

Preliminary results of the US Naval Observatory Twin Astrograph
Catalog (TAC) project show FDP's up to about $200 mas = 2 \mu$
(in preparation for publication).
The FDP's of the blue and yellow lenses
are different, although the plates have been measured on the
same machine.

\subsection{Physical explanation for the FDP}

The {\em cause} of these field distortions is
certainly not the reference star catalog,
because the FDP is a systematic error with respect to
exactly the size of a plate.
Candidates for an explanation of the FDP are
effects in the camera lens, the filter, the photographic plate,
the darkroom technique, and the entire plate measuring process.
The bending of the Schmidt plates, causing
a FDP of very large amplitude, is a different issue and
will not be discussed here.

We will now investigate the effect of the plate measuring process
on the FDP.
Assume, there is a calibration error of the measuring machine, $c(x_{m},
y_{m})$,
depending on the machine coordinates $x_{m}, y_{m}$.
Let a plate be measured in 2 orientations, the plate being rotated
by exactly $180\deg$ between the 2 sets of measures,
with $x_{m} = y_{m} = 0$ as rotation center,
then the residual calibration error $rc(x_{m}, y_{m})$,
after combining the 2 sets of measures of the plate is

\[ rc (x_{m}, y_{m}) = ( c (x_{m}, y_{m}) - c (-x_{m}, -y_{m}) ) / 2  \]

which is anti--symmetric, i.e. $rc(x,y) = -rc(-x,-y)$.
This holds for any (small) calibration error pattern of a
measuring machine as long as it is the same for both sets of measures.
The CPC2 plates have been measured exactly this way.

The FDP of Fig.1 was split into an anti--symmetric part,
shown in Fig.8a, and a remainder, shown in Fig.8b.
The pattern in Fig.8a might, but need not be caused by the measuring process,
while the pattern of Fig.8b can not be introduced by the measuring
process with the assumptions made above.

Note also, that the camera was rotated by $90\deg$ with respect
to the sky after the Cape Zone plates were taken (Paper I).
All plates were observed with the camera on one side of
the telescope mounting pier.
Assuming the FDP is caused by the camera lens,
one would expect the FDP from the Cape Zone plates to be rotated by $90\deg$
with respect to the FDP of the other data.
This is not the case.
Contrary, the magnitude equation in photographic
astrometry can be explained by decentering of the lens
(Platais et al. 1994).

The small magnitude equation previously found
in the CPC2 data (Paper II) is partly consistent with the assumption
of being caused by the lens.
A magnitude equation in $x$ was found in the Cape Zone, while
in the zone $\delta \ge -30\deg$ it is in $y$, consistent with
the rotation of the lens by $90\deg$.
But there is no significant magnitude equation found in the
polar zone, which had been observed simultaneously with the
equator zone.

The position of the filter with respect to the sky or telescope
and possible changes of the position of the filter during the course
of the CPC2 observations are unknown.

In a different investigation,
the FDP of Fig.1 was split up in an east--west symmetric
part (Fig.9a) and a remainder (Fig.9b).
The mean absolute vector lengths are given in Table 1.
The east--west symmetric part contains nearly all information,
while the non--symmetric part is dominated by noise.
There is no proven explanation for this nearly perfect
east--west symmetry at the moment.
The drying process of the plates in the darkroom
is one possibility.
Assume most of the plates were standing upright with the
same orientation (e.g. north up) while drying.
Gravity could cause a north--south
(up--down) asymmetry, while the plate boundaries
lead to an east--west (right--left) symmetry.

\subsection{Outlook}

No unique conclusion about the physical origin of the FDP in the CPC2
can be reached so far. Possible causes are inhomogeneities in the
filter used to obtain the spectral bandpass or
the plate development process in the darkroom.
The Cape Astrograph has been out of operation for more than 20 years now
and no further tests can be made.
Considering the small systematic errors we are investigating here,
a large statistical set of data is required anyway for testing.

The US Naval Observatory (USNO) Twin Astrograph yellow lens
was used in a northern hemishpere sky catalog,
as well as in work for the extragalactic
reference link in the southern hemisphere.
The former plates were measured at the USNO and the latter are
being measured at the Hamburg Observatory.
Evaluation of both sets of data will give some more insight
into this issue, specifically a separation of effects
depending on the measuring machine and other effects will be possible.

Once the Hipparcos Output Catalog becomes available,
a much more precise evaluation of all systematic errors of the
photographic material can be made.
The Hipparcos Output Catalog will contain about 2.5 times more reference
stars per plate than the currently available
reference star catalogs, and its limiting magnitude
is fainter by about 2 magnitudes.
The average random error of the Hipparcos Ouput Catalog star positions
at the mean epoch of the CPC2 is expected to be only $\le 50 mas$.
The correlation between the pure geometric FDP and possible
magnitude and coma terms will then be investigated, as well
as possible variations in the optical distortion term on a
plate by plate basis, which is not possible with
the SRS catalog.

It is advisable to {\em inspect all photographic data}
with respect to these field distortions.
Various methods can be applied.
Simple binning of residuals is appropriate when enough
reference stars are available.
Future wide field CCD imaging of overlapping areas
in the sky should be investigated similarly
(Zacharias et al. 1995).
This will also give some information about the
distinction between effects related to photographic plates
and other effects causing the FDP.

\section{Conclusions}

Small systematic errors in the CPC2 plate data (x,y measures)
as a function of the location on the plate (field distortion pattern)
were found with a maximum vector length of $160 mas = 1.6 \mu m$
and an average of 57 mas.
The FDP is nearly independent of the area in the sky observed, and
thus, independent of the zenith distance.
It is also not dependent on the magnitudes of the stars and is very stable
with time (epoch or season of observation).
The FDP can be corrected at the $0.1 \mu m$ level, which
will improve the precision of the individual observations.
The mean catalog position is nearly unaffected by
the FDP due to the favorable 4--fold overlap observing strategy
used in the CPC2 project, but this need not to be the case
for other projects.

The {\em actual variation} of the field distortions
from {\em plate to plate} can be determined only with a
very dense, high precision reference star catalog, which is
not available now.
In all patterns shown here, only the {\em average} mean
field distortion from a large number of plates can be determined,
which hopefully is the largest contribution to the total field
distortion of a particular plate.
The large number of plates in the CPC2 project, however, allows
spliting up the data into a reasonable number of subsets,
which has made possible this detailed investigation of the
variation of the FDP.

Use of the 2--dimensional FDP is a more appropriate and more
general method for dealing with the purely geometric, systematic
errors in photographic astrometry
(when the statistical sample is large enough)
than is use of only 1--dimensional projections as
radial residual versus radius.

Applying corrections for the field
distortions to the x,y coordinates prior to plate adjustment
is mandatory for a {\em block adjustment} of that
plate material.
Without the FDP corrections applied, the BA positions
were found to be biased by up to 100 mas at the edge of the
covered sky area.
Thus, a BA is very sensitive to systematic errors in the photographic
data and can easily lead to biased positions as shown above.

It has been decided not to publish the improved CPC2 as obtained
by this investigation because the Hipparcos Output Catalog will soon
become available for a much better investigation of the
systematic errors in the CPC2 plate data, which will ultimately
lead to an unbiased BA result.

Important for future photographic (and CCD) surveys is to keep constant
the orientations of the plates (the CCD) and filter in all processing steps.
This will allow the systematic errors to be well determined and
the data could then be corrected for FDP with sufficient accuracy.

\begin{table}
\caption{Properties of field distortion patterns (FDP's) from various subsets
          of the data.
       The first four columns give reference to the figure numbers in this
       paper, the subset selection criterium, the number of plates and
       residuals used within the subset, respectively.
 	 $\overline{x}$ = arithmetic mean    of x--component of residuals;
 	 $\sigma_{x}$   = standard deviation of x--component of residuals;
 	 $\overline{y}$, $\sigma_{y}$ = similar for y--component;
 	 $\overline{l}$, $\sigma_{l}$ = arithmetic mean
	   and standard deviation of vector length of residuals;
 	 $\sigma_{\Delta{x}}$, $\sigma_{\Delta{y}}$, $\sigma_{\Delta{l}}$
 			= standard deviation of the difference in residuals
 			  (subset FDP minus FDP of all data) for the x--component,
 			  the y--component and the vector length, respectively.}

\end{table}

\newpage

\begin{figure}
\caption{Field Distortion Pattern (FDP) generated from all CPC2 data.
  The field size is 4.07\deg, north is up and east is to the right.
  The scale of the residual vectors is increased by 10,000.}
\end{figure}

\begin{figure}
\caption{Radial residuals vs. distance of star from the plate center (radius).
   a) before applying the FDP, b) after applying the FDP.}
\end{figure}

\begin{figure}
\caption{a) The FDP for plates centered at declination
       $-52\deg \le \delta \le -40\deg$ (Cape Zone),
	 b) the difference of a) minus the mean FDP of all zones as of Fig.1.}
\end{figure}

\begin{figure}
\caption{Precision of field star positions in declination ($\sigma_{\delta}$)
	 from the conventional plate adjustment of overlapping plates
	 as function of magnitude. The open squares are derived from a solution
	 of the original x,y data and the filled circles are derived from the
	 solution with x,y data corrected for FDP as of Fig.1.}
\end{figure}

\begin{figure}
\caption{Combined FDP of a plate area in the sky covered by a
	 regular 4--fold overlap.}
\end{figure}

\begin{figure}
\caption{Block Adjustment (BA) minus Conventional Plate Adjustment (CPA)
	 declinations of all stars vs. declination,
	 a) before b) after correction for FDP.
	 Each point is the average of 1,000 differences.}
\end{figure}

\begin{figure}
\caption{The same as Fig.5 but for a 2--fold overlap along right ascension,
	 which is the plate pattern in the most northern $2\deg$ of the CPC2.}
\end{figure}

\begin{figure}
\caption{The FDP pattern of all data as of Fig.1 has been split up into two
parts,
	 a) the anti-symmetric part (f(x,y) = -f(-x,-y)) and b) the remainder.}
\end{figure}

\begin{figure}
\caption{The FDP pattern of all data as of Fig.1 has been split up into two
parts,
	 a) the east-west-symmetric part (f(x,y) = -f(-x,y)) and b) the remainder.}
\end{figure}

\end{document}